\documentclass[twocolumn,floatfix,showpacs,nofootinbib,10pt,groupedaddress,superscriptaddress,]{revtex4-1}
\usepackage{amssymb,amsmath,graphicx}
\usepackage{color}
\usepackage[colorlinks,hyperindex]{hyperref}
\usepackage{slashed}
\hypersetup{hidelinks,backref=true,pagebackref=true,hyperindex=true,colorlinks=true,breaklinks=true,urlcolor= blue}
\hypersetup{%
  colorlinks = true, 
  linkcolor  = blue
}
\usepackage{caption}
\usepackage{multirow}
\usepackage{comment}
\usepackage{slashbox}
\usepackage{bookmark}
\usepackage{hyperref,color,xcolor}
\hypersetup{hidelinks,backref=true,pagebackref=true,hyperindex=true,colorlinks=true,breaklinks=true,urlcolor= blue}
\hypersetup{%
  colorlinks = true,
  linkcolor  = blue
}
\definecolor{green1}{RGB}{0,128,0} 
\hypersetup{hidelinks,backref=true,pagebackref=true,hyperindex=true,colorlinks=true,breaklinks=true,urlcolor= blue}
\hypersetup{%
  colorlinks = true,
  linkcolor  = blue,
  citecolor = green1,
}

\usepackage{bookmark,textgreek}
\usepackage{hyperref,color,xcolor}
\hypersetup{hidelinks,hyperindex=true,colorlinks=true,breaklinks=true,urlcolor= blue}
\hypersetup{%
  colorlinks = true,
  linkcolor  = blue
}
\usepackage{makeidx,upgreek}
\newcommand{\beq}{\begin{eqnarray}}
\newcommand{\benu}{\begin{enumerate}}
\newcommand{\enu}{\end{enumerate}}
\newcommand{\eeq}{\end{eqnarray}}
\newcommand{\be}{\begin{equation}}
\newcommand{\ee}{\end{equation}}

\newcommand{\ba}{\begin{eqnarray}}
\newcommand{\ea}{\end{eqnarray}}
\begin{document}
\title{The Heisenberg spinor field classification and its interplay with the Lounesto's classification}
\author{Marcos R. A. Arcod\'ia}
\email{marcodia@mdp.edu.ar}
\affiliation{Instituto de Investigaciones F\'isicas de Mar del Plata (IFIMAR), Consejo Nacional de Investigaciones
Cient\'ificas y T\'ecnicas (CONICET), Mar del Plata, Argentina.}
\author{Mauricio Bellini}
\email{mbellini@mdp.edu.ar}
\affiliation{Departamento de F\'isica, Facultad de Ciencias Exactas y Naturales, Universidad Nacional de Mar
del Plata, Funes 3350, C.P. 7600, Mar del Plata, Argentina}
\affiliation{Instituto de Investigaciones F\'isicas de Mar del Plata (IFIMAR), Consejo Nacional de Investigaciones
Cient\'ificas y T\'ecnicas (CONICET), Mar del Plata, Argentina.}
\author{Rold\~ao~da~Rocha}
\email{roldao.rocha@ufabc.edu.br}
\affiliation{Federal University of ABC, Center of Mathematics, Computing and Cognition, 09210-580, Santo Andr\'e, Brazil}\email{roldao.rocha@ufabc.edu.br}

\begin{abstract}
Dirac linear spinor fields were obtained from non-linear Heisenberg spinors, in the literature. Here we extend that idea by considering not only Dirac spinor fields but spinor fields in any of the Lounesto's classes. When one starts considering all these classes of fields, the question of providing a classification for the Heisenberg spinor naturally arises. In this work the classification of Heisenberg spinor fields is derived and scrutinized, in its interplay with the Lounesto's spinor field classification.

\end{abstract}
\maketitle
\section{Introduction}
The so-called Lounesto's  classification of spinor fields is constituted of six classes of regular and singular spinor fields, assorted 
with respect to the values attained by their respective bilinear covariants. The Majorana, Dirac, and Weyl spinors, although occupying a 
privileged spot in this classification, are discrete points in six immense spinor spaces. Several non-standard spinor fields, beyond the Majorana, Dirac, and Weyl ones, were found and studied in this context. Refs. \cite{Cavalcanti:2014wia,Fabbri:2016msm} 
pave a reciprocal Lounesto's classification, asserting the most general type of spinor field in each one of Lounesto's class. 
Important samples of new spinor fields can be seen, e. g., in Refs. \cite{exotic,esk,Ablamowicz:2014rpa,Dantas:2015mfi,Vaz:2017fac,Fabbri:2011mi,Fabbri:2014foa,Ahluwalia:2009rh,Rogerio:2017gvr,HoffdaSilva:2016ffx,Rogerio:2016mxi,HoffdaSilva:2017waf,HoffdaSilva:2009is}. In addition, new classifications in string theory \cite{BBR} and  new anyonic spinor fields \cite{Lopes:2018cvu} were derived and scrutinized.
Once the Lounesto's standard spinor field classification is restricted to the U(1) gauge symmetry, splitting off classes of charged and neutral spinors, Ref. \cite{Fabbri:2017lvu} then implemented other gauge symmetries, where in particular the gauge symmetry in electroweak theory was introduced and spinor doublets were then classified and studied. The Lounesto's  spinor field classification corresponds, in this extended classification,  to a sole Pauli singlet \cite{Fabbri:2017lvu}. 
VSR symmetries and DKP algebras were also investigated in the context of Lounesto spinor fields in Refs. \cite{Cavalcanti:2014uta,Robinson:2018exx}. 

On the other hand, very little is known about the Heisenberg spinors in this context. 
The Heisenberg equation governs the dynamics of diverse spinor fields, constituting the Inomata--McKinley spinor fields one of its particular solution. On the other hand, some regular spinor fields can be constructed upon appropriate linear mixtures of Inomata--McKinley spinor fields \cite{Beghetto:2017nmb}.
Dirac spinor fields were described by a non-linear mixture of Heisenberg spinors fields  \cite{Novello:2007cb}, to show that neutrinos are quantum field states of Heisenberg spinor fields. 
Besides, Heisenberg dynamics was employed to study anisotropic cosmological models, generated by a non-linear fermionic ultra-relativistic fluid \cite{Joffily:2016klu}.

The main aim here is to propose a spinor field classification of 
Heisenberg spinor fields, and relate it to the Lounesto's classification.  A byproduct of our development in this paper is, in particular, to emulate previous constructions that describe Dirac spinor fields as Heisenberg ones, to further encompass all types of spinor fields in both the Heisenberg and the Lounesto's classification. 
This paper is organized as follows: after reviewing the Lounesto's classification and presenting the bilinear covariants, the regular and singular spinor fields in Sect. \ref{II}, Sect. \ref{III} devotes to derive and present the Heisenberg classification of spinor fields and to 
scrutinize the interplay between Heisenberg spinor fields and the ones in the Lounesto's classification. Sect. \ref{c} 
draws the concluding remarks and further discussion. 

\section{Lounesto's classification and ramifications}
\label{II}
Spinors in the  Minkowski spacetime, $M$,  are elements pertaining to the spinor bundle on $M$, being carrier spaces of the so-called ${\left(1/2, 0\right)}\oplus{\left(0, 1/2\right)}$ representations of the Lorentz group.  
For their huge spectrum of applications, it is sometimes better to work with arbitrary bases $\{\upgamma^\mu\}$ of the $\Omega(M)=\oplus_{i=0}^4\Omega^i(M)$ exterior bundle. Bilinear covariants consist of homogeneous sections  of the exterior bundle \cite{Cra}, 
\begin{widetext}
\begin{center}
\begin{tabular}[c]{|c| c| c| c| c|}
\hline
$\;\;\Upomega^0(M)\;\;$&$\;\;\Upomega^1(M)\;\;$&$\;\;\Upomega^2(M)\;\;$&$\;\;\Upomega^3(M)\;\;$&$\;\;\Upomega^4(M)\;\;$
\\\hline
 $\textcolor{black}{\upsigma}$ & $\mathbb{J}=J_{\mu}\upgamma^{\mu}$ & $\;\;\mathbb{S}=S_{\mu\nu }\upgamma^{\mu}\wedge \upgamma^{ \nu }\;\;$&$\mathbb{K}= K_{\mu }\upgamma^{\nu }$&$\textcolor{black}{\upomega}$\\
\hline
\end{tabular}\captionof{table}{Bilinear covariants as homogeneous sections of the spin bundle.}\label{bilinearcov}
\end{center} 
\end{widetext}
where 
\begin{subequations}
\begin{eqnarray}
\textcolor{black}{\upsigma}&=&\textcolor{black}{\bar{\psi}\psi},\label{upsig}\\
K_{\mu }&=&\bar{\psi}\upgamma_{5}\upgamma _{\mu }\psi,\label{jj}\\
S_{\mu\nu}&=&\frac{i}{2}\bar{\psi}[\upgamma _{\mu},\upgamma_{
\nu }]\psi,\label{ss}\\
J_{\mu}&=&\bar{\psi}\upgamma _{\mu }\psi,\label{kk}\\
\textcolor{black}{\upomega}&=&\textcolor{black}{i\bar{\psi}\upgamma_{5}\psi\,,}\label{upom}
\end{eqnarray}
\end{subequations}
 are coefficients of Lorentz bilinear covariants. The Clifford algebra definition of generators,  $\upgamma_{\mu }\upgamma _{\nu
}+\upgamma _{\nu }\upgamma_{\mu }=2\eta_{\mu \nu }\mathbf{1}$ is also assumed. In addition, denoting the Clifford product by juxtaposition, the 
 volume element, $\upgamma_5=i\upgamma_0\upgamma_1\upgamma_2\upgamma_3$, implements the chiral operator. To fix the notation, the conjugation $\bar\psi=\psi^\dagger\upgamma_0$ will be regularly used. 

The Lounesto's classification consists of the following six classes of spinor fields:
\begin{subequations}
\begin{eqnarray}
&&(1)\;\;\;\mathbb{S}\neq 0, \;\;\;\mathbb{K}\neq0,\;\;\;\upsigma\neq0,\;\;\; \upomega\neq0,\;\;\text{}
\label{tipo1}\\
&&(2)\;\;\;\mathbb{S}\neq 0, \;\;\;\mathbb{K}\neq0,\;\;\;\upsigma=0,\;\;\;  \upomega\neq0,\;\;\label{tipo2}\\   
&&(3)\;\;\;\mathbb{S}\neq 0, \;\;\;\mathbb{K}\neq0,\;\;\;\upsigma\neq0,\;\;\;  \upomega=0,\label{tipo3}\\  
&&(4)\;\;\;\mathbb{S}\neq 0, \;\;\;\mathbb{K}\neq0,\;\;\;\upsigma=0=\upomega,  \;\;\text{}\quad\qquad\label{tipo4}\\
&&(5) \;\;\;\mathbb{S}\neq 0, \;\;\;\mathbb{K}=0,\;\;\;\upsigma=0=\upomega,\text{}\quad\qquad\label{tipo5}\\
&&(6)\;\;\;\mathbb{S}=0, \;\;\;\mathbb{K}\neq0,\;\;\;\upsigma=0=\upomega.\;\;\;\text{}\quad\qquad\label{tipo6}
\end{eqnarray}
\end{subequations}
\textcolor{black}{Physical observables, as the current density and spin density in the electron Dirac's theory, correspond to the bilinear covariants $\mathbb{J}$ and $\mathbb{K}$, respectively, whereas $\mathbb{S}$ plays the role of the spin density. More precisely, after an appropriate scaling by the electron charge, $e$, the Planck constant (over 2$\pi$), $\hbar$, and the speed of light, $c$, the temporal component $eJ_0$ is interpreted as the electrical charge density, whereas $ec J_k$ ($i,j,k=1,2,3$) represents the electric current density. 
The spatial components $(e\hbar/2mc) S^{ij}$ stand for the magnetic moment density, and the $(e\hbar/2mc) S^{0j}$ is the electric moment density. Finally $(\hbar/2) K_\mu$ is the chiral current density. The interpretation of the scalar, $\sigma$, and pseudoscalar, $\omega$, is less clear, however  $\sigma^2+\omega^2$ can be interpreted as a probability density. In addition, $\sigma$ appears as mass and self-interaction terms in  spinor Lagrangians, whereas $\omega$, being CP-odd, might probe CP features. }. However, not all the spinor fields in the Lounesto's classification share similar interpretations. In addition, for some other particular subclasses of the Lounesto's classification, the Fierz--Pauli--Kofink relations can be verified
\cite{Cra}: 
\begin{subequations}
\begin{eqnarray}\label{fifi1}
\upsigma S_{\alpha\beta}\epsilon_{\;\;\;\rho\sigma}^{\alpha\beta}-\upomega{S}_{\rho\sigma}&=&\epsilon_{\rho\sigma\alpha\beta}J^\rho {K}^\sigma,\\\eta_{\rho\sigma}(J^\rho J^\sigma+K^\rho K^\sigma)&=&0=\eta_{\rho\sigma}J^\rho K^\sigma,\label{fifi2}\\ \upsigma^{2}+\upomega^{2}&=&\eta_{\rho\sigma}J^\rho J^\sigma\,.\label{fifi3}  
\end{eqnarray}
\end{subequations}
\noindent  
The inequality $\mathbb{J}\neq0$ is valid for the entire classes in Lounesto's classification.  Ref. \cite{EPJC} constructed three more classes that are beyond Lounesto's classification, corresponding to $\mathbb{J}=0$, playing the role of ghost spinor fields. 
 
The Lounesto's classification splits the spinor fields into singular ones, with both vanishing $\upsigma$ and $\upomega$, and regular ones, where at least one between the scalar and pseudoscalar bilinear covariants are not equal to zero. Dirac spinor fields are regular spinor fields of type-(1), whereas Majorana ones are in a subclass of type-(5), flagpole,  spinor fields. Besides, Weyl spinor fields are dipole spinor fields of type-(6). Singular spinor fields have, besides the Majorana and Weyl spinors, other representatives  \cite{Ahluwalia:2009rh,daRocha:2005ti,Alves:2014kta,Alves:2017joy}. Most of the subclasses in the Lounesto's classification have been explored and a lot of room is available for spinor fields that can play important roles in high energy physics \cite{esk}.

 The Fierz--Pauli--Kofink relations, (\ref{fifi1}) -  (\ref{fifi3}), are not, in general, satisfied by all the singular spinor fields. The very definition of a Fierz aggregate, as a multivector field whose homogeneous components are the bilinear covariants themselves, 
 \begin{eqnarray}
\mathbb{Z}= \upsigma-(\upomega+\mathbb{K})\upgamma_{5}+ \mathbb{J}+i\mathbb{S}  \,, \label{Z}\label{zsigma}
\end{eqnarray} yields the relations 
 \begin{subequations}
\begin{eqnarray}\label{zz}
&&-4iK_{\mu}\mathbb{Z}=\mathbb{Z}\upgamma_{5}\upgamma_{\mu}\mathbb{Z}\\
&&-4i\upomega\mathbb{Z}= \mathbb{Z}\upgamma_{5}\mathbb{Z},\\
&&-4iS_{\mu\nu}\mathbb{Z}=\mathbb{Z}\upgamma_{\mu}\upgamma_{\nu}\mathbb{Z},\\
&&-4iJ_{\mu}\mathbb{Z}= \mathbb{Z}\upgamma_{\mu}\mathbb{Z}.\label{zzz}
\,
\end{eqnarray}
It replaces the Fierz--Pauli--Kofink relations (\ref{fifi1}) -(\ref{fifi3}) for the singular spinor fields that do not obey them.
Eqs. (\ref{zz}) - (\ref{zzz}) are the most general ones and hold for any spinor field in the entire Lounesto's classification. 

\end{subequations}

\section{Heisenberg spinor fields classification}
In what follows natural units shall be used. 
\label{III}
Let one considers the Heisenberg equation \cite{Novello:2007cb}
\begin{equation}
i\upgamma^{\mu} \partial_{\mu} \, \mathring\psi - 2 s \, (\mathring\upsigma + i\mathring\upomega
\upgamma^{5}) \, \mathring\psi = 0 \label{heiseq}
\end{equation}for a Heisenberg
spinor field, $\mathring\psi$, 
in which the constant $s$ has the dimension of (length)$^2$. Hereon ringed quantities are constructed with respect to  Heisenberg spinors. The Heisenberg scalar and pseudoscalar bilinear covariants respectively read
\begin{eqnarray}
\mathring\upsigma&=& \overline{\mathring\psi} \mathring\psi\,, \label{ringsigma}\\
\mathring\upomega &=& i \overline{\mathring\psi}\upgamma^{5} \mathring\psi\,.
\label{ringomega}
\end{eqnarray}
Hereon the spinor field $\uppsi$, which may lie in any Lounesto spinor field class, will be named a \emph{Lounesto spinor field}, whereas $\mathring{\psi}$ will be called a Heisenberg spinor field.  
Let $\mathring{\mathbb{J}}$, $\mathring{\mathbb{S}}$, $\mathring{\mathbb{K}}$  be the Heisenberg bilinear covariants corresponding to $\mathring{\psi}$, with definitions analogous to (\ref{jj}) - (\ref{kk}). By using Eq. (\ref{exp10}), it is possible to compute the bilinear covariants of the $\uppsi$ Lounesto spinor field in terms of the ones corresponding to the $\mathring{\psi}$ Heisenberg spinor field.
Defining the left and right chiral projectors,
\begin{equation}\label{projectors}
P_{L}:=\frac{1}{2}(1+\upgamma^{5}) \ , \ P_{R}:=\frac{1}{2}(1-\upgamma^{5}),
\end{equation}
one can write $\mathring{\psi}_{L}=P_{L}\mathring{\psi}$ and $\mathring{\psi}_{R}=P_{R}\mathring{\psi}$, as well as ${\uppsi}_{L}=P_{L}{\uppsi}$ and ${\uppsi}_{R}=P_{R}{\uppsi}$.
The Heisenberg spinor $\mathring\psi$ can be split into chiral ones 
\begin{equation}
\psi = \mathring\psi_{R}+ \mathring\psi_{L} = \frac{1}{2} (1 -
\upgamma^{5})\mathring\psi + \frac{1}{2} (1 + \upgamma^{5})\mathring\psi.
\label{projs}
\end{equation}
\textcolor{black}{Given $F,G$ complex numbers,} writing an arbitrary spinor field in the Lounesto's classification as  
\begin{equation}
\uppsi =  e^{G} \, \mathring\psi_{R} + e^{F} \, \mathring\psi_{L} 
\label{exp10}
\end{equation}
is equivalent of emulating the usual expression for chiral spinor fields, $\uppsi_{L} = e^{F} \, \mathring\psi_{L}$ and
$\uppsi_{R} = e^{G} \, \mathring\psi_{R}.$ \textcolor{black}{The physical role played by $F$ and $G$ and useful expressions involving the dynamics of Heisenberg and Inomata spinors are presented in Appendix \ref{app}}. 

Ref. \cite{Novello:2007cb} imposes some conditions on a Dirac spinor field, for it to be described as a particular mixture of Heisenberg spinor fields.  Let one remembers that a Dirac spinor field is a very particular state into 
the first class of spinors in Lounesto's classification.  Now we extend this idea and consider not only Dirac spinor fields (\ref{tipo1}) but the entire Lounesto's classification of spinor fields, (\ref{tipo1}) - (\ref{tipo6}). 
In order to implement such an extension, it will be necessary to first provide a classification of Heisenberg spinor fields and 
the interplay between the Lounesto's classes of spinor fields and Heisenberg spinor fields.

To relate the standard bilinear covariants to the \textcolor{black}{ones}  constructed upon Heisenberg spinor fields, the scalar $\upsigma=\bar{\uppsi}\uppsi$ is written as
\begin{eqnarray}
\upsigma&=&(e^{F}P_{L}\mathring{\psi}+e^{G}P_{R}\mathring{\psi})^{\dagger}\upgamma^{0}(e^{F}P_{L}\mathring{\psi}+e^{G}P_{R}\mathring{\psi})\nonumber\\
&=&{\mathring{\psi}}^{\dagger}(e^{F^{*}}P_{L}^{\dagger}+e^{G^{*}}P_{R}^{\dagger})\upgamma^{0}(e^{F}P_{L}+e^{G}P_{R})\mathring{\psi}.
\end{eqnarray} 
Using the fact that ${P_{R}}$ and ${P_{L}}$ are self-adjoint, together with $P_{L}\upgamma^{0}=\upgamma^{0}P_{R}$ and $P_{R}\upgamma^{0}=\upgamma^{0}P_{L}$, yields 
\begin{equation}
\upsigma=\overline{\mathring{\psi}}(e^{F^{*}}P_{R}+e^{G^{*}}P_{L})(e^{F}P_{L}+e^{G}P_{R})\mathring{\psi}.
\end{equation}
In addition, as  $P_{R}$ and $P_{L}$ are orthogonal  idempotents, Eqs. (\ref{projectors}) imply that 
\begin{eqnarray}
\!\!\!\!\!\!\!\!\!\!\!\!\upsigma&=&\overline{\mathring{\psi}}(e^{F+G^{*}}P_{L}+e^{G+F^{*}}P_{R})\mathring{\psi}\\
\!\!\!\!\!\!\!\!\!\!\!\!&=&\overline{\mathring{\psi}}\Bigg(\frac{e^{F+G^{*}}+e^{G+F^{*}}}{2}+\frac{e^{F+G^{*}}-e^{G+F^{*}}}{2}\upgamma^{5}\Bigg)\mathring{\psi}.
\end{eqnarray} 
Defining the complex number $z=e^{F+G^{*}}$ yields 
\begin{equation}
\upsigma=\overline{\mathring{\psi}}(\text{Re}(z)+\text{Im}(z)i\upgamma^{5})\mathring{\psi}=\text{Re}(z)\mathring{\upsigma}+\text{Im}(z)\mathring{\upomega}
\end{equation}
Performing similar calculations for the other bilinear covariants, the following set of equations, that sets the bilinear covariants with respect to the Heisenberg ones, can be obtained, 
\begin{subequations}
\begin{eqnarray}
&&\upsigma=\text{Re}(z)\mathring{\upsigma}+\text{Im}(z)\mathring{\upomega}\\
&&J^{\mu}=\frac{y+x}{2}\mathring{J}^{\mu}+\frac{y-x}{2}\mathring{K}^{\mu},\\
&&S^{\mu\nu}=\text{Re}(z)\mathring{S}^{\mu\nu}-\text{Im}(z)\star\mathring{S}^{\mu\nu},\\
&&K^{\mu}=\frac{y-x}{2}\mathring{J}^{\mu}+\frac{y+x}{2}\mathring{K}^{\mu},\\
&&\upomega=\text{Re}(z)\mathring{\upomega}-\text{Im}(z)\mathring{\upsigma},
\end{eqnarray}
\end{subequations}
where $x=e^{F+F^{*}}=e^{2\text{Re}(F)}$ and $y=e^{G+G^{*}}=e^{2\text{Re}(G)}$. This system of equations is not fully coupled, being equivalent to three independent systems of equations, 
\begin{eqnarray}\label{scalars}
\begin{pmatrix}
	\upsigma \\
    	\upomega \\
\end{pmatrix}
&=&
\begin{pmatrix}
   \text{Re}(z) & \text{Im}(z)  \\
  -\text{Im}(z) & \text{Re}(z) \\
\end{pmatrix}
\begin{pmatrix}
	\mathring \upsigma \\
    	\mathring \upomega \\
\end{pmatrix},\\\label{currents}
\begin{pmatrix}
	{J}^{\mu} \\
    	{K}^{\nu} \\
\end{pmatrix}
&=&\frac{1}{2}
\begin{pmatrix}
   y+x & y-x \\
   y-x & y+x \\
\end{pmatrix}
\begin{pmatrix}
	\mathring{J}^{\mu} \\
    	\mathring{K}^{\mu} \\
\end{pmatrix},\\\label{bivector}
\begin{pmatrix}
	{S}^{\mu\nu} \\
    	\star{S}^{\mu\nu} \\
\end{pmatrix}
&=&
\begin{pmatrix}
   \text{Re}(z) & -\text{Im}(z)  \\
   \text{Im}(z) & \text{Re}(z) \\
\end{pmatrix}
\begin{pmatrix}
	\mathring{S}^{\mu\nu} \\
    	\star\mathring{S}^{\mu\nu} \\
\end{pmatrix},
\end{eqnarray}
where $\star$ denotes the Hodge operator, that satisfies $\star\star=-id$ in $M$. Also,  $i\overline{\uppsi}\upgamma^{\mu}\upgamma^{\nu}\upgamma^{5}\uppsi=-{\star{S}}^{\mu\nu}$ and $i\overline{\mathring{\psi}}\upgamma^{\mu}\upgamma^{\nu}\upgamma^{5}\mathring{\psi}=-{\star{\mathring{S}}}^{\mu\nu}$.
Since the modulus of the complex number $z\neq0$ is always positive, then the matrices in (\ref{scalars}) and (\ref{bivector}) are invertible. Furthermore, since the determinant of the matrix in (\ref{currents}) is $xy/2$ and $x\neq{0}$ and $y\neq{0}$, this matrix is also invertible. In order to establish the classification of the Heisenberg spinor, it will be helpful to invert these equations, arriving at:
\begin{equation}\label{scalarsinv}
\begin{pmatrix}
	\mathring \upsigma \\
    	\mathring \upomega \\
\end{pmatrix}
=\frac{1}{|z|^{2}}
\begin{pmatrix}
   \text{Re}(z) & -\text{Im}(z)  \\
   \text{Im}(z) & \text{Re}(z) \\
\end{pmatrix}
\begin{pmatrix}
	\upsigma \\
    	\upomega \\
\end{pmatrix},
\end{equation}
\begin{equation}\label{currentsinv}
\begin{pmatrix}
	\mathring{J}^{\mu} \\
    	\mathring{K}^{\mu} \\
\end{pmatrix}
=\frac{1}{2xy}
\begin{pmatrix}
   y+x & x-y \\
   x-y & y+x \\
\end{pmatrix}
\begin{pmatrix}
	{J}^{\mu} \\
    	{K}^{\nu} \\
\end{pmatrix},
\end{equation}
\begin{equation}\label{bivectorinv}
\begin{pmatrix}
	\mathring{S}^{\mu\nu} \\
    	\star\mathring{S}^{\mu\nu} \\
\end{pmatrix}
=\frac{1}{|z|^{2}}
\begin{pmatrix}
   \text{Re}(z) & \text{Im}(z)  \\
   -\text{Im}(z) & \text{Re}(z) \\
\end{pmatrix}
\begin{pmatrix}
	{S}^{\mu\nu} \\
    	\star{S}^{\mu\nu} \\
\end{pmatrix}.
\end{equation}
Aiming to establish a classification for the Heisenberg spinor $\mathring{\psi}$, analogous to the Lounesto classification for the field $\uppsi$, at a first sight one could think of a classification for $\mathring\psi$ with the exact same classes proposed by Lounesto. 
\textcolor{black}{However, Eq. (\ref{currentsinv}) shows that ${\mathbb{J}}\neq{0}$ does not imply $\mathring{\mathbb{J}}\neq{0}$, and since all the Lounesto classes assume non-vanishing current density, the Heisenberg spinor field classification has to contain more classes in order to allow $\mathring{\mathbf{J}}=0$.}

Due to these facts, the  Heisenberg spinor field classification is contained within the following classes:
\begin{subequations}
\begin{eqnarray}
&&(\mathring{1})\ \;\;\;\mathbb{\mathring{K}}\neq 0, \;\;\;\mathbb{\mathring{S}}\neq0,\;\;\;\mathring{\upomega}\neq0,\;\;\;  \mathring{\upsigma}\neq0,\;\;\text{}
\label{tipo1heis}\\
&&(\mathring{2})\ \;\;\;\mathbb{\mathring{K}}\neq 0, \;\;\;\mathbb{\mathring{S}}\neq0,\;\;\;\mathring\upomega=0,\;\;\;  \mathring\upsigma\neq0,\;\;\label{tipo2heis}\\   
&&(\mathring{3})\ \;\;\;\mathbb{\mathring{K}}\neq 0, \;\;\;\mathbb{\mathring{S}}\neq0,\;\;\;\mathring\upomega\neq0,\;\;\;  \mathring\upsigma=0,\label{tipo3heis}\\  
&&(\mathring{4})\ \;\;\;\mathbb{\mathring{K}}\neq 0, \;\;\;\mathbb{\mathring{S}}\neq0,\;\;\;\mathring\upomega=0=\mathring\upsigma,  \;\;\text{}\quad\qquad\label{tipo4heis}\\
&&({\mathring{4}}^{*})\;\;\;\mathbb{\mathring{K}}\neq 0, \;\;\;\mathbb{\mathring{S}}\neq0,\;\;\;\mathring\upomega=0=\mathring\upsigma,\;\;\; \mathbb{\mathring{J}}=0,\;\;\; \text{}\quad\qquad\label{tipo4heisnull}\\
&&(\mathring{5})\ \;\;\;\mathbb{\mathring{K}}= 0, \;\;\;\mathbb{\mathring{S}}\neq0,\;\;\;\mathring\upomega=0=\mathring\upsigma,\text{}\quad\qquad\label{tipo5heis}\\
&&({\mathring{5}}^{*}) \;\;\;\mathbb{\mathring{K}}= 0, \;\;\;\mathbb{\mathring{S}}\neq0,\;\;\;\mathring\upomega=0=\mathring\upsigma, \;\;\; \mathbb{\mathring{J}}=0, \text{}\quad\qquad\label{tipo5heisnull}\\
&&(\mathring{6})\ \;\;\;\mathbb{\mathring{K}}\neq 0, \;\;\;\mathbb{\mathring{S}}=0,\;\;\;\mathring\upomega=0=\mathring\upsigma,\;\;\;\text{}\quad\qquad\label{tipo6heis}\\
&&({\mathring{6}}^{*})\;\;\;\mathbb{\mathring{K}}\neq 0, \;\;\;\mathbb{\mathring{S}}=0,\;\;\;\mathring\upomega=0=\mathring\upsigma, \;\;\; \mathbb{\mathring{J}}=0. \;\;\;\text{}\quad\qquad\label{tipo6heisnull}
\end{eqnarray}
\end{subequations}

In this list $\mathbb{\mathring{J}}\neq{0}$ unless the contrary is stated. We have used the same numbering as in the Lounesto classification but adding a ``$*$'' superscript to the classes with $\mathbb{\mathring{J}}=0$. Note that by virtue of Eq. (\ref{fifi3}), this is only possible for singular spinors. 
From Eqs. (\ref{scalarsinv}) -- (\ref{bivectorinv}), it is clear that a spinor field is Lounesto-singular if and only if it is Heisenberg-singular. Hence, classes (\ref{tipo1}), (\ref{tipo2}) and (\ref{tipo3}) remain within classes (\ref{tipo1heis}), (\ref{tipo2heis}) and (\ref{tipo3heis}); and  classes (\ref{tipo4}), (\ref{tipo5}) and (\ref{tipo6}) remain within classes (\ref{tipo4heis}), (\ref{tipo4heisnull}), (\ref{tipo4heis}), (\ref{tipo5heisnull}), (\ref{tipo6heis}) and (\ref{tipo6heisnull}).
Moreover, the Heisenberg spinor field cannot belong to the class (\ref{tipo5heisnull}), for $\mathbb{\mathring{J}}={0}$ and $\mathbb{\mathring{K}}=0$ imply $\mathbb{K}={0}=\mathbb{J}$, but $\mathbb{J}\neq{0}$ must always hold. 

In what follows every case in the Lounesto's spinor field classification for $\uppsi$ will be studied, deriving the corresponding Heisenberg spinor fields $\mathring{\psi}$ that  give origin to it.
 \medbreak
\underline{Type-1 Lounesto spinor field:} in this case, as asserted before, $\mathring{\psi}$ does correspond to a regular spinor field. Hence $\mathring\upsigma\neq{0}$ or $\mathring\upomega\neq{0}$. Hence $\mathbb{\mathring{J}}\neq{0}$, as stated above. This implies that $\mathbb{\mathring{K}}\neq{0}$ as well, for it being zero would contradict the Fierz identity (\ref{fifi3}), once one uses (\ref{fifi2}). 
Also, Eq. (\ref{bivectorinv}) yields $\mathbb{\mathring{S}}\neq{0}$.
Eq. (\ref{scalarsinv}) implies that 
\begin{equation}
\begin{gathered}
\mathring{\upsigma}=\frac{\text{Re}(z)\upsigma-\text{Im}(z)\upomega}{|z|^{2}},\\
\mathring{\upomega}=\frac{\text{Im}(z)\upsigma+\text{Re}(z)\upomega}{|z|^{2}}.
\end{gathered}
\end{equation}
It is evident that in the case of $z$ being either real or imaginary, then $\mathring{\psi}$ is of type (\ref{tipo1heis}). If $z$ is neither real nor a pure imaginary, then different possibilities arise. 

If the condition $\upsigma=-\frac{\text{Re}(z)}{\text{Im}(z)}\upomega$ is satisfied, then one has $\mathring{\upsigma}\neq{0}$ and $\mathring\upomega={0}$. Consequently $\mathring{\psi}$ is of type (\ref{tipo2heis}). 
The condition $\upsigma=\frac{\text{Im}(z)}{\text{Re}(z)}\upomega$ yields $\mathring{\upsigma}=0$ and $\mathring\upomega\neq{0}$, and hence the Heisenberg spinor belongs to the class (\ref{tipo3heis}).
If neither of these conditions hold, then $\mathring\upsigma\neq{0}$ and $\mathring\upomega\neq{0}$, and in consequence $\mathring{\psi}$ is an element of the class (\ref{tipo1heis}).
\medbreak
\underline{Type-2 Lounesto spinor:} $\mathring{\psi}$ will also be regular. Hence $\mathring\upsigma\neq{0}$ or $\mathring\upomega\neq{0}$.
\medbreak
As explained for the two previous types, $\mathbb{\mathring{J}}\neq{0}$, $\mathbb{\mathring{K}}\neq{0}$ and $\mathbb{\mathring{S}}\neq{0}$. Eq. (\ref{scalarsinv}) yields
\begin{equation}
\begin{gathered}
\mathring{\upsigma}=\frac{-\text{Im}(z)}{|z|^{2}}\upomega, \\
\mathring{\upomega}=\frac{\text{Re}(z)}{|z|^{2}}\upomega.
\end{gathered}
\end{equation}
Hence, if both $\text{Re}(z)$ and $\text{Im}(z)$ are non null, then $\mathring{\psi}$ is of type (\ref{tipo1heis}). If $z$ is real, then $\mathring{\psi}$ is in the class (\ref{tipo3heis}). If $z$ is a pure imaginary, then the Heisenberg spinor field lies into the class (\ref{tipo2heis}).
\medbreak
\underline{Type-3 Lounesto spinor field:} Analogously, the spinor field $\mathring{\psi}$ will be regular. Hence, either $\mathring\upsigma\neq{0}$ or $\mathring\upomega\neq{0}$. Similarly for the Heisenberg type-1 case, $\mathbb{\mathring{J}}\neq{0}$, $\mathbb{\mathring{K}}\neq{0}$ and $\mathbb{\mathring{S}}\neq{0}$. Eq. (\ref{scalarsinv}) yields
\begin{equation}
\begin{gathered}
\mathring{\upsigma}=\frac{\text{Re}(z)}{|z|^{2}}\upsigma, \\
\mathring{\upomega}=\frac{\text{Im}(z)}{|z|^{2}}\upsigma.
\end{gathered}
\end{equation}
Hence, if both $\text{Re}(z)$ and $\text{Im}(z)$ are different of zero, then $\mathring{\psi}$ is of type (\ref{tipo1heis}). If $z$ is real, then $\mathring{\psi}$ is in the class (\ref{tipo2heis}), and if $z$ is a pure imaginary, then the Heisenberg spinor field is an element of the class (\ref{tipo3heis}).
\medbreak
\underline{Type-4 Lounesto spinor field:} this spinor field is singular, hence $\mathring{\upsigma}=\mathring{\upomega}=0$. Since $\mathbb{S}\neq{0}$, then $\mathbb{\mathring{S}}\neq{0}$. The vector and pseudovector bilinear covariants,  for the Heisenberg field,  are given by
\begin{equation}\label{curr}
\begin{gathered}
\mathbb{\mathring{J}}=\frac{(y+x)\mathbb{J}+(x-y)\mathbb{K}}{2xy} \ , \\
\mathbb{\mathring{K}}=\frac{(x-y)\mathbb{J}+(x+y)\mathbb{K}}{2xy} \ .
\end{gathered}
\end{equation}
If $x=y$ then $\mathbb{\mathring{J}}=\mathbb{J}/{x}\neq{0}$, $\mathbb{\mathring{K}}=\mathbb{K}/{x}\neq{0}$. In this case $\mathring{\psi}$ belongs to the class (\ref{tipo4heis}).

If $x\neq{y}$ then we have three possibilities. If the condition $\mathbb{J}=\frac{y-x}{y+x}\mathbb{K}$ is satisfied, then it can be seen from (\ref{curr}), that $\mathbb{\mathring{J}}=0$ and $\mathbb{\mathring{K}}\neq{0}$. Hence, the Heisenberg spinor lies in the class (\ref{tipo4heisnull}). 
If the condition $\mathbb{J}=\frac{y+x}{y-x}\mathbb{K}$ is satisfied, it is clear from (\ref{curr}) that $\mathbb{\mathring{J}}\neq{0}$ and $\mathbb{\mathring{K}}={0}$. Hence, the Heisenberg spinor is of type (\ref{tipo5heis}).
If neither of the former two conditions hold, then $\mathbb{\mathring{J}}\neq{0}$ and $\mathbb{\mathring{K}}\neq{0}$ and, in consequence,  $\mathring{\psi}$ is of type (\ref{tipo4heis}).
\medbreak
\underline{Type-5 Lounesto spinor field:} as in the former case, this spinor field is singular, as $\mathring{\upsigma}=\mathring{\upomega}=0$. Since $\mathbb{S}\neq{0}$, then $\mathbb{\mathring{S}}\neq{0}$ and Eq. (\ref{currentsinv}) yields
\begin{equation}
\begin{gathered}
\mathbb{\mathring{J}}=\frac{(y+x)\mathbb{J}}{2xy} \ , \\
\mathbb{\mathring{K}}=\frac{(x-y)\mathbb{J}}{2xy} \ .
\end{gathered}
\end{equation}
If $x=y$ then $\mathbb{\mathring{J}}\neq{0}$ and $\mathbb{\mathring{K}}={0}$. In this case $\mathring{\psi}$ belongs to the class (\ref{tipo5heis}). If $x\neq{y}$, then we have that $\mathbb{\mathring{J}}\neq{0}$ and $\mathbb{\mathring{K}}\neq{0}$. In consequence, the Heisenberg spinor field is of class (\ref{tipo4heis}).
\medbreak
\underline{Type-6 Lounesto spinor:} This case is very similar to the type-4 one, as $\mathring{\upsigma}=\mathring{\upomega}=0$ and $\mathbb{\mathring{S}}={0}$. 
The vector and pseudovector equations for the Heisenberg field are the same that in the type-4 spinor (\ref{curr}). Hence, if $x=y$ then $\mathbb{\mathring{J}}=\mathbb{J}/{x}\neq{0}$, $\mathbb{\mathring{K}}=\mathbb{K}/{x}\neq{0}$. In this case $\mathring{\psi}$ belongs to the class (\ref{tipo6heis}).

If $x\neq{y}$ then three possibilities arise.
If the condition $\mathbb{J}=\frac{y-x}{y+x}\mathbb{K}$ is satisfied, then it can be seen from (\ref{curr}) that $\mathbb{\mathring{J}}=0$ and $\mathbb{\mathring{K}}\neq{0}$. The Heisenberg spinor field, in this case, lies in the class (\ref{tipo6heisnull}).
If the condition $\mathbb{J}=\frac{y+x}{y-x}\mathbb{K}$ holds, then (\ref{curr}) yields $\mathbb{\mathring{J}}\neq{0}$ and $\mathbb{\mathring{K}}={0}$. Hence, the Heisenberg spinor is of type (\ref{tipo5heis}).
If neither of the former conditions hold, then $\mathbb{\mathring{J}}\neq{0}$ and $\mathbb{\mathring{K}}\neq{0}$ and in consequence $\mathring{\psi}$ is of type (\ref{tipo6heis}).
\medbreak
One can combine all these results in the following tables for regular and singular Lounesto spinor fields:

\begin{center}
\begin{tabular}{|c|c|c|c|c|}
\hline
\multicolumn{5}{|c|}{Regular Spinors}
\\ \hline
\multicolumn{2}{|l|}{\backslashbox{Conditions \ \ \ \ }{Lounesto Spinors}}                                                           & 1              & 2                               & 3                               \\ \hline
\multirow{3}{*}{$z$ not pure} & $\upsigma=\frac{\text{Im}(z)}{\text{Re}(z)}\upomega$  & $\mathring{3}$ & \multirow{3}{*}{$\mathring{1}$} & \multirow{3}{*}{$\mathring{1}$} \\ \cline{2-3}
                          & $\upsigma=-\frac{\text{Re}(z)}{\text{Im}(z)}\upomega$ & $\mathring{2}$ &                                 &                                 \\ \cline{2-3}
                       & None of the above                                 & $\mathring{1}$ &                                 &                                 \\ \hline
\multicolumn{2}{|c|}{$z$ real} & \multicolumn{1}{c|}{$\mathring{1}$} & \multicolumn{1}{c|}{$\mathring{2}$} & \multicolumn{1}{c|}{$\mathring{3}$} \\ \hline
\multicolumn{2}{|c|}{$z$ imaginary} & \multicolumn{1}{c|}{$\mathring{1}$} & \multicolumn{1}{c|}{$\mathring{3}$} & \multicolumn{1}{c|}{$\mathring{2}$} \\
\hline
\end{tabular}
\captionof{table}{Correspondence between regular classes.}\label{tableregular}
\end{center}
\begin{center}
\begin{tabular}{|c|c|c|c|c|}
\hline
\multicolumn{5}{|c|}{Singular Spinors}
\\ \hline
\multicolumn{2}{|l|}{\backslashbox{Conditions}{Lounesto Spinors}}                                                           & 4              & 5                               & 6                               \\ \hline
\multirow{3}{*}{$x\neq{y}$} & $\mathbb{J}=\frac{y-x}{y+x}\mathbb{K}$  & $\mathring{4}^{*}$ & \multirow{3}{*}{$\mathring{4}$} & $\mathring{6}^{*}$ \\ \cline{2-3}  \cline{5-5}
                          & $\mathbb{J}=\frac{y+x}{y-x}\mathbb{K}$ & $\mathring{5}$ &                                 & $\mathring{5}$ \\ \cline{2-3} \cline{5-5}
                      & None of the above                                 & $\mathring{4}$ &                                 &  $\mathring{6}$  \\ \hline
\multicolumn{2}{|c|}{$x=y$} & \multicolumn{1}{c|}{$\mathring{4}$} & \multicolumn{1}{c|}{$\mathring{5}$} & \multicolumn{1}{c|}{$\mathring{6}$} \\ \hline
\end{tabular}
\captionof{table}{Correspondence between singular classes.}\label{tablesingular}
\end{center} 

Observe that Tables \ref{tableregular} and \ref{tablesingular} are expressed using the auxiliar parameters $x$, $y$ and $z$, which depend upon the functions $F$ and $G$. Using 
\begin{equation}
\begin{gathered}
x=\exp(2\text{Re}(F))\ ,\\
y=\exp(2\text{Re}(G))\ ,\\
z=\exp(F+G^{*})\ ,
\end{gathered}
\end{equation}
and the following equivalences 
\begin{equation}\nonumber
\begin{gathered}
x=y \text{ if and only if } \text{Re}(F)=\text{Re}(G),\\
z \text{ is real iff } \text{Im}(F)-\text{Im}(G)\equiv{0} \pmod \pi ,\\
z \text{ is imaginary iff } \text{Im}(F)-\text{Im}(G)\equiv{\pi/2}\pmod \pi.
\end{gathered}
\end{equation}
then the results in the tables can be expressed using $F$ and $G$ instead of $x$, $y$ and $z$.

The results shown in the tables comprise the interplay between the Lounesto's and the Heisenberg spinor fields, 
yielding a new classification that can play a prominent role in searching for new fermionic fields and their applications 
in gravitation and field theory. Paving the Heisenberg spinor field also makes the Lounesto's classifications to be better understood. 

\section{Conclusions}\label{c}

We have considered a spinor field contained in any class of the Lounesto's classification as a particular linear combination of right and left projections of a Heisenberg spinor field. In that scenario we aimed to establish a classification for the Heisenberg spinor field and a correspondence between the two classifications. Such a goal is attained, the results being expressed in Tables \ref{tableregular} and \ref{tablesingular}.

Some interesting points about the interplay between Heisenberg and Lounesto spinor are observed. First, a spinor in a Lounesto's class is regular if and only if it comes from a regular Heisenberg spinor, which allowed us to split the results into Tables \ref{tableregular} and \ref{tablesingular}. Second, from these tables, one can realize that the correspondence between regular classes depend only on the imaginary part of the functions $F$ and $G$,  in Eq. (\ref{exp10}), whereas for singular spinors it depends on the real part of these functions. Last, provided that $\uppsi$ is a Lounesto spinor field, the corresponding Heisenberg spinor field can never belong to the class (\ref{tipo5heisnull}). Hence, not all the classes with vanishing current density in Ref. \cite{EPJC} are necessary for the Heisenberg field classification.

Let one considers an arbitrary spinor field, $\uppsi$, whose dynamics is governed by the Dirac equation. It is worth to emphasize that the fact that $\uppsi$ satisfies the Dirac equation 
does not impose  almost any condition on $\uppsi$, in the context of the Lounesto's classification.
In fact, there are regular and singular spinor fields, in every single class in the Lounesto's classification
that satisfies the Dirac equation. Hence, although the Dirac equation imposes some type of spinor in each class,
the amount of spinor fields in each class in Lounesto's classification is still large.

\textcolor{black}{
One of our motivations to introduce the Heisenberg spinor field classification was to better understand the Lounesto's classification, as well as to provide a complementary point of view that has been still not encompassed by the Lounesto's classification. The main motivation regarding this classification consists of the dynamics and kinematics that governs the spinor fields in all subclasses in Lounesto's classification. For example, Majorana (mass dimension 3/2) and  Elko spinor fields (mass dimension 1) \cite{Ahluwalia:2009rh,Ahluwalia:2008xi,horv1} are neutral spinor fields in the class 5 of flagpoles spinor fields \cite{daRocha:2005ti}, according to the Lounesto's classification. However, there is a charged spinor field, representative of a subclass in the class 5 of Lounesto's classification, that satisfies the Dirac equation in some gravitational background \cite{pan}. Hence, only knowing the class in Lounesto's classification does not guarantee to find the dynamics ruling the spinor field, although some exclusive options are clear once the equations of motion are known. As the Heisenberg classification, here proposed, is somehow related to the Heisenberg equation, it can therefore provide a clue on the dynamics of the related spinor field in Lounesto's classification.  
As important as the Wigner classification in this context is the classification in Ref. \cite{Fabbri:2016msm}, as it places a physical classification of the degrees of freedom of the spinor fields. Complementary equations to the Fierz--Pauli--Kofink identities were shown to be similar to the Pauli--Lubanski axial vector, and the classification in  Ref. \cite{Fabbri:2016msm} emulates the Wigner one. We can just concretely assert about resemblances between the Heisenberg and the Wigner classification after deriving and exploring the analogue of  the Fierz--Pauli--Kofink identities in the Heisenberg classification. As they might be obtained from the Fierz--Pauli--Kofink identities (\ref{fifi1}) -- (\ref{fifi3}), since the Heisenberg and Lounesto spinors are related by Eq. (\ref{exp10}), it is worth in a near future to implement the  analogue of  the Fierz--Pauli--Kofink identities for the Heisenberg classification, to precisely answer this important question.}

In Ref. \cite{Novello:2007cb} a Dirac spinor (class 1 in the Lounesto's classification) was built up from a Heisenberg spinor, and solutions to the Dirac equation were obtained using Inomata--McKinley solutions of the Heisenberg equation. Tables \ref{tableregular} and \ref{tablesingular} would make possible to seek for solutions of the Dirac equation for other classes of Lounesto spinor fields. As a prototypical Heisenberg dynamics was employed to study anisotropic cosmological models \cite{Joffily:2016klu}, one may then explore this  kind of dynamics for other types of Heisenberg spinor fields in the context of the minimal geometric deformation \cite{Casadio:2016aum}. 

\acknowledgments
MB and MA acknowledge CONICET, Argentina (PIP 11220150100072CO) and UNMdP
(EXA852/18), for financial support.
MA thanks to UFABC, for the hospitality. 
RdR~is grateful to CNPq (Grant No. 303293/2015-2),
and to FAPESP (Grant No. 2017/18897-8), for partial financial support. 
\appendix
\textcolor{black}{\section{Inomata solutions of the Heisenberg equation}
\label{app}
A particular solution of the Heisenberg  equation can be derived 
when one considers the Inomata condition 
\begin{equation}
\partial_{\mu} \Psi = \left( a \, J_{\mu} + b\, K_{\mu}
\gamma^{5} \right)  \,\Psi \label{inom}
\end{equation}
for $a,b\in \mathbb{C}$. 
Any $\Psi$ spinor field that satisfies Eq. (\ref{inom}) is an 
Inomata spinor field. The integrability condition of Eq. (\ref{inom}) reads 
$
 Re (a) = Re (b)$.
Let $R,S$ be scalar fields such that $
K_{\mu} = \partial_{\mu} R$ and $J_{\mu} = \partial_{\mu} S$, 
or equivalently \cite{Novello:2007cb}
\begin{eqnarray}
\!\!\!\!\!\!\!\!\!\!\!\!\!\!\!R = \frac{1}{b - \overline{b}}  \log \left( \frac{ \mathring\upsigma - i
\mathring\upomega}{\sqrt{J_{\mu} J^{\mu}}} \right),\quad S  = \frac{1}{a+ \overline{a}}   \log \sqrt{J_{\mu} J^{\mu}}.
\protect\label{sss}
\end{eqnarray}
Ref. \cite{Novello:2007cb} then shows the following respective expressions for $F$ and $G$:
\begin{eqnarray}
\!\!\!\!\!\!\!\!\!\!\!\!\!\!\!F &=& \!-\frac{1}{2} (b \!-\! \overline{b} ) R + \left[2is -
\frac{1}{2}(b - \overline{b})\right] S 
\!+\! \frac{ i M} {a \!+\!
\overline{a}}
 \, e^{-(a \!+\! \overline{a}) S}\\
\!\!\!\!\!\!\!\!\!\!\!\!\!\!\!G &=&\! \frac{1}{2}  (b \!-\! \overline{b} )R + \left[ 2 i s \!-\!
\frac{1}{2} (b \!-\! \overline{b})\right] S \!+\! \frac{ i M} {a \!+\!
\overline{a}} e^{-(a \!+\! \overline{a}) S}.
\end{eqnarray}}

\end{document}